\documentclass[12pt,bm,verbatim,graphicx,amsmath,wasysym]{article}
\usepackage[T1]{fontenc}
\usepackage{newtxmath,newtxtext}
\usepackage{graphicx}  % needed for figures
\usepackage{url}
\usepackage[dvipsnames]{xcolor}
\usepackage{authblk}

\title{Monopole Excitation and Nuclear Compressibility: Present and Future Perspectives}

\author[1]{J.~C. Zamora}
\author[1]{S. Giraud}
\affil[1]{Facility for Rare Isotope Beams, Michigan State University, East Lansing, Michigan 48824, USA}

\date{}

\usepackage[backend=biber, uniquelist=false, style=apa]{biblatex}
\usepackage{hyperref}
\hypersetup{hidelinks}

\begin{document}

\maketitle

\begin{abstract}

Isoscalar giant resonances are nuclear collective excitations associated with the oscillation in phase of protons and neutrons according to a certain multipolarity $L$. In particular, the isoscalar giant monopole resonance ($L=0$) is the strongest nuclear compression mode, and its excitation energy is directly related to the  compression modulus  for finite nuclei.  Typically, microscopic calculations are utilized to establish a relationship between the experimental compression modulus and the nuclear incompressibility that is a crucial parameter of the equation of state for nuclear matter. The incompressibility of nuclear matter has been determined with an accuracy of 10 to 20\% using relativistic and non-relativistic microscopic models for describing the monopole distributions in ${}^{208}$Pb and ${}^{90}$Zr isotopes.  However, the same theoretical models are not able to describe  data for open-shell nuclei, such as those of tin and cadmium isotopes. In fact, only effective interactions with a softer nuclear-matter incompressibility are able to predict the centroid energy of monopole distributions for open-shell nuclei. An unified description of the monopole resonance in ${}^{208}$Pb and other open-shell nuclei remains unsolved from the theory side. Most of this uncertainty is due to our poor knowledge of the symmetry energy, which is another essential component of the equation of state of nuclear matter. Therefore, new experimental data along isotopic chains covering a wide range in $N/Z$  ratios, including neutron-deficient and neutron-rich nuclei, are of paramount importance for determining both the nuclear-matter incompressibility and the symmetry energy more precisely.  Novel approaches using inverse kinematics have been developed to achieve giant resonance experiments with unstable nuclei. The active target and storage ring are potentially the most feasible methods for measuring giant resonances in nuclei far from stability. In the near future, the combination of these techniques with high-intensity radioactive beams at new accelerator facilities will provide the means to explore the nuclear-matter properties of the most exotic nuclei.

\end{abstract}

\section{Introduction}
 
 One of the major challenges of contemporary astrophysics and nuclear physics is describing the behavior of matter under extreme conditions. Understanding the structure and dynamics of superdense objects in the universe, such as neutron stars, is one of the most compelling open problems. In recent decades, considerable efforts have been made to increase our understanding of these intriguing scenarios. Recent astronomical observations of an unusual gamma-ray burst and the discovery of gravitational waves originating from a binary neutron-star merger have significantly advanced the study of dense matter physics \parencite{Abbott2017}. In fact, these observations marked the beginning of the multi-messenger era, which combines various techniques such as neutrino detection, X-ray telescopes, radio astronomy, etc \parencite{Neronov_2019}.  All of these multi-messenger signals provide essential information about these cosmic objects. \par
 
 Terrestrial experiments are also  fundamental to constraint the theoretical models that are employed to describe these  phenomena. The atomic nucleus provides  ideal conditions for simulating stellar media and extracting matter properties in the laboratory. The atomic mass is concentrated in the nucleus, which is a few femtometers (fm) in size and has a saturation density of approximately $\rho_0 \approx 2.7\times 10^{14}$~g/cm$^3$. However, a neutron star has a radius of approximately 10~km, but its interior matter density can be an order of magnitude higher than the nuclear saturation density ($\rho \approx 10 \rho_0$) \parencite{LATTIMER2007109}. Nuclear reaction experiments at intermediate and high energies provide the means to compress nuclear matter to such high densities and extrapolate to the stellar conditions. \par
  
\begin{figure}[!ht]
\centering
\includegraphics[width=0.8\textwidth]{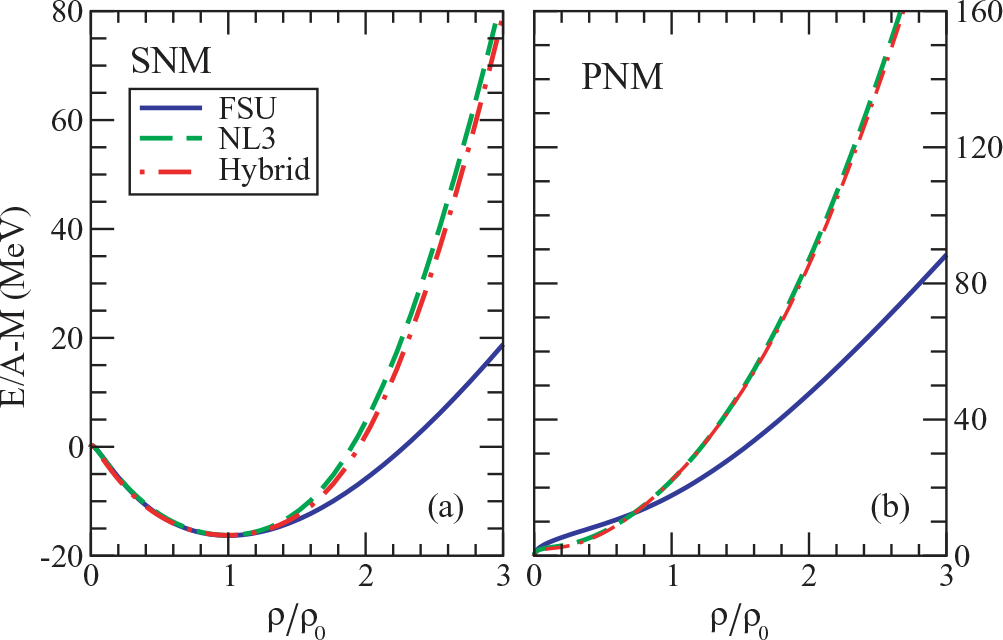}
\caption{\label{fig1} Density dependence of the energy per nucleon ($E/A-M \equiv {\cal E}(\rho, \delta)$) in symmetric nuclear matter (left) and in pure neutron matter (right) using three different models.   The density is expressed as a ratio to the nuclear saturation density. Adapted from \parencite{PhysRevC.79.054311}. }
\end{figure}

Although the atomic nucleus is nearly 20 orders of magnitude smaller than a neutron star, the equation of state (EoS) of nuclear matter connects these two objects \parencite{sym13030400}.  The EoS is a thermodynamic equation that relates state variables (such as pressure, temperature, density, energy, isospin, etc.) that characterize nuclear matter under a particular set of physical conditions. In the thermodynamic limit, nuclear matter is idealised as a Fermi fluid composed of an infinite number of protons and neutrons interacting via the nuclear force. At zero temperature, the total energy per nucleon, ${\cal E}(\rho, \delta)$, is characterized by the total nucleonic density $\rho$ and the isospin asymmetry $\delta = (\rho_n-\rho_p)/\rho$ [$\rho_n$($\rho_p$) is the neutron (proton) density].  ${\cal E}(\rho, \delta)$ can be expanded to the second order as:
 \begin{equation}
  {\cal E}(\rho, \delta) = {\cal E}_\text{SMN}(\rho) + S(\rho)\delta^2,
  \label{eq1}
 \end{equation}
where ${\cal E}_\text{SMN}$ is the specific energy in the symmetric nuclear matter (SNM), and $S$ is the symmetry energy. Note that the first order of the expansion is not included because odd powers of $\delta$ do not contribute due to the symmetry of the strong force between like-nucleon pairs \parencite{PhysRevC.79.054311}. The leading term of this expansion, ${\cal E}_\text{SMN}(\rho)$, corresponds to the contribution from symmetric matter (i.e., $\delta=0$).  The second-order correction to the symmetric limit is the symmetry energy, which quantifies the amount of energy required to convert part of the protons in symmetric nuclear matter to excess neutrons (pure neutron matter, $\delta=1$). Figure~\ref{fig1} illustrates the total energy per nucleon ($E/A$) relative to the nucleon rest mass ($M$)  as a function of  density  using different effective interactions for both symmetric nuclear matter (SNM) and pure neutron matter (PNM).  FSUGold and NL3 are well-known models that successfully reproduce the ground-state properties of nuclei, while the Hybrid model was built to reproduce ``softer" matter conditions \parencite{PhysRevC.79.054311}.

Around the saturation density ($\rho_0$), the symmetric nuclear matter can be expanded in a Taylor series up to third order, with a conveniently defined dimensionless parameter $x= (\rho-\rho_0)/3\rho_0$, as

\begin{equation}
 {\cal E}_\text{SMN}(\rho) = {\cal E}_0 + \frac{1}{2}K_\text{nm} x^2 + \frac{1}{6}Q_0 x^3,
 \label{eq2}
\end{equation}
where ${\cal E}_0$ is the energy per particle of symmetric nuclear matter at $\rho_0$, $K_\text{nm}$ is the incompressibility and $Q_0$ is the skewness parameter of symmetric nuclear matter. Similarly, the symmetry energy  is expanded around the saturation density up to third order
\begin{equation}
 S(\rho) = J + Lx + \frac{1}{2}K_\text{sym} x^2 + \frac{1}{6}Q_\text{sym} x^3,
 \label{eq3}
\end{equation}
where $J$, $L$, $K_\text{sym}$ and $Q_\text{sym}$ are the values of the symmetry energy, slope, curvature and skewness, respectively. \par

The parameters of both symmetric and symmetry matter (Eqs.~(\ref{eq2}) and (\ref{eq3})) are constrained from experiments, particularly from isoscalar giant resonances.  Nuclear compression modes like isoscalar giant monopole
and dipole resonances are of great interest because
their energies are directly related to the nuclear-matter
incompressibility.\\
Although studies on giant resonances have been going on for many years, there are still many unanswered questions, especially regarding the matter incompressibility and collective modes in unstable nuclei, which may have a significant impact  constraining the nuclear equation of state. An overview of the current status of research in isoscalar giant resonances, still-unresolved problems, and potential future directions are presented in this work.

\section{Nuclear compression modes}

\begin{figure}[!ht]
\centering
\includegraphics[width=0.8\textwidth]{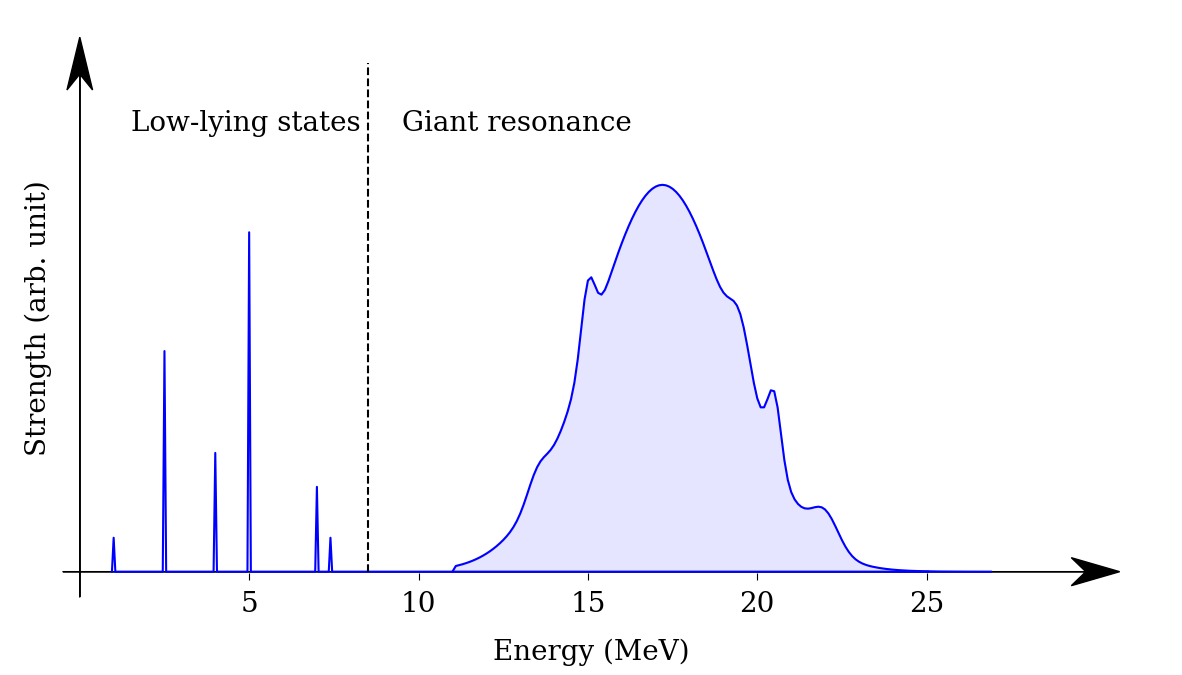}
\caption{\label{fig2} Representation of the cross section as a function of excitation energy. Below the particle separation energy (dashed line), the nucleus responds through the excitation of simple states involving only one or a few particles. At higher excitation energies, the nucleus responds with a collective motion involving all the nucleons which results in a broad resonance from approximately 10 to 30~MeV.  }
\end{figure}

The atomic nucleus is a many-body quantum system comprised of protons and neutrons that are confined in a very small space ($\sim$~fm). Many-body systems in nature, such as the atomic nucleus \parencite{harakeh2001}, but also in biology \parencite{Karsenti2008}, classical mechanics \parencite{STROGATZ20000}, condensed matter \parencite{IGNESMULLOL2007163}, quantum information \parencite{Lewis-Swan2019}, and many others, exhibit self-organization phenomena. In particular, protons and neutrons in the atomic nucleus can exhibit a self-organization mechanism due to the effect of a driving force (nuclear excitation), where most, if not all, of the nucleons participate in a collective nuclear motion. Such collective nuclear modes are known as giant resonances. Giant resonances can be considered as high-frequency ($10^{21}$~Hz) damped vibrations around the equilibrium shape of the nucleus. The restoring force of the resonance is directly related to the bulk properties of the nucleus, such as its compression modulus. Giant resonances are excited at energies above the particle separation threshold ($\sim 8$ to 10~MeV).
Figure~\ref{fig2} illustrates a typical cross section as a function of the excitation energy, which is  characterized by low-lying states and a giant resonance. Giant resonances are usually concentrated at excitation energies between 10 and 30~MeV with a relatively large cross section, close to the maximum allowed by sum-rule arguments \parencite{STRINGARI1982232}. As any other resonance, giant resonances  are characterized by their centroid, width, and strength, which are directly related to the ground-state properties of the excited nucleus. \par

\begin{figure}[!ht]
\centering
\includegraphics[width=0.75\textwidth]{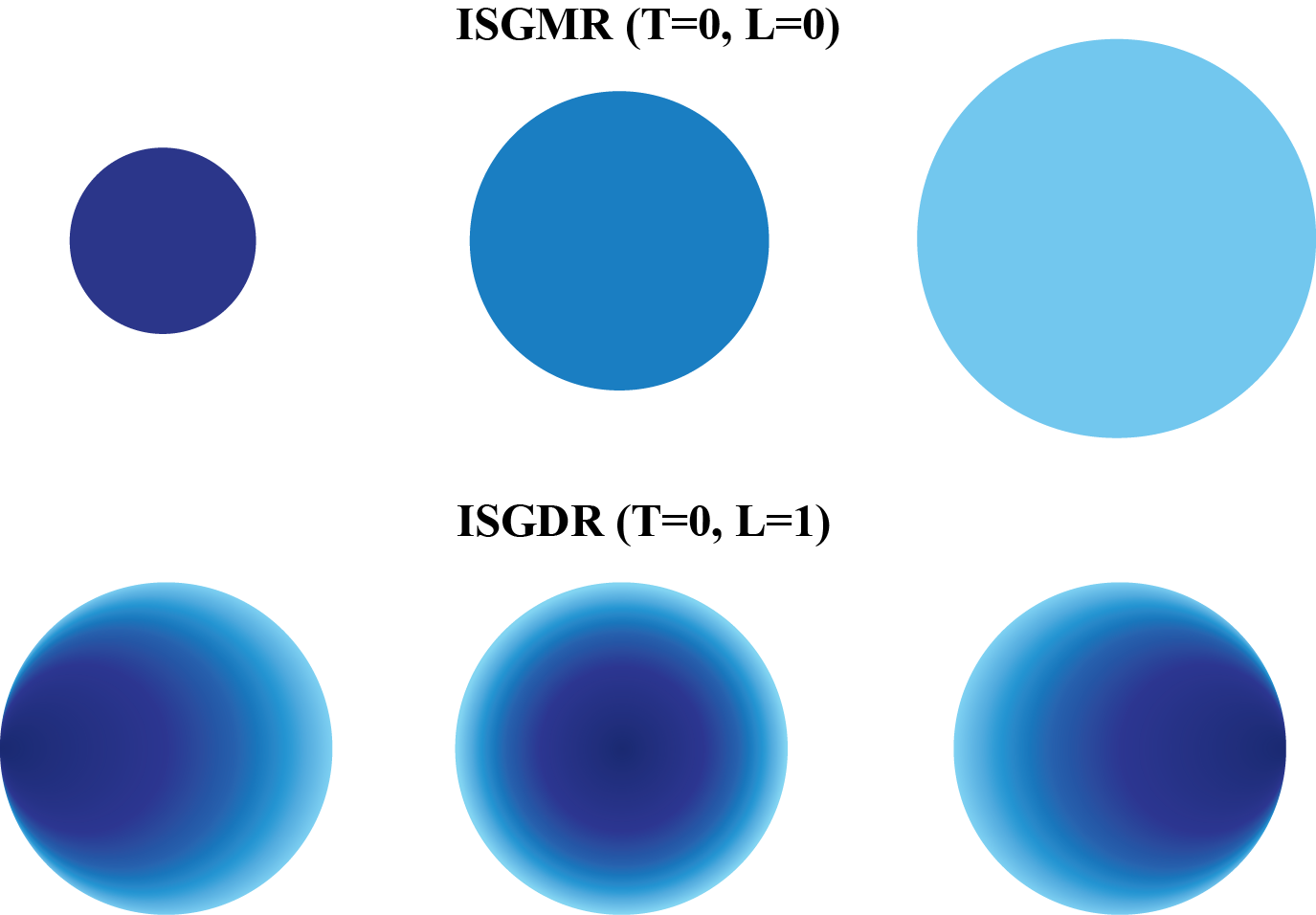}
\caption{\label{fig3} Representation of the nuclear compression modes ISGMR (isoscalar giant monopole resonance) and ISGDR (isoscalar giant dipole resonance). Protons and neutrons move in phase in the isoscalar modes. The ISGMR can be assumed as a radial oscillation of the nucleus in a "breathing" mode. The ISGDR is an overtone ($3\hbar\omega$) excitation associated with a density oscillation similar to a "squeezing" mode.  }
\end{figure}

Giant resonances are classified depending on their multipolarity ($L$) and the isospin ($T$) and spin ($S$) degrees of freedom of the nucleons \parencite{harakeh2001}. In particular, the $\Delta S=0$, $\Delta T=0$ modes are known as the isoscalar giant vibrations that are associated with the oscillation in phase of protons and neutrons according to a  multipolarity pattern given by  $\Delta L$. For example, $\Delta L=0$ corresponds to the isoscalar giant monopole resonance (ISGMR), which in a macroscopic picture  can be assumed to be a radial oscillation of the nucleus in a "breathing" mode. The multipolarity $\Delta L=1$ ($3\hbar\omega$) also corresponds to a compression mode, the isoscalar giant dipole resonance (ISGDR), which can be understood as a nuclear density oscillation, often called as  "squeezing" mode. Figure~\ref{fig3} shows a representation of these compression modes.

\begin{figure}[!ht]
\centering
\includegraphics[width=0.5\textwidth]{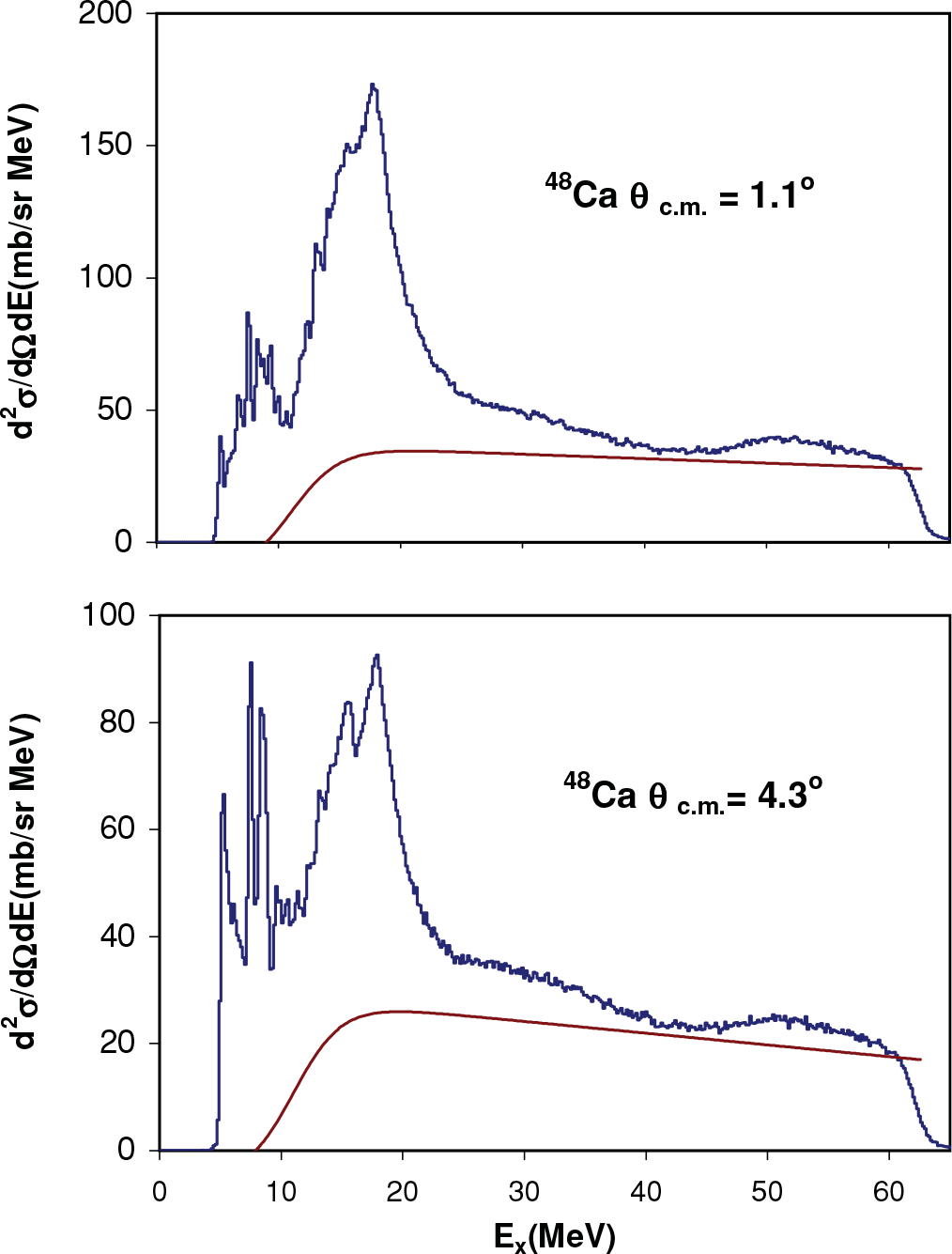}
\caption{\label{fig4} Example of excitation energy spectra measured from inelastic $\alpha$ scattering on ${}^{48}$Ca. The top and bottom figures correspond to measurements at center-of-mass angles of 1.1 and 4.3$^\circ$, respectively.  Adapted from \parencite{PhysRevC.83.044327} }
\end{figure}

Inelastic scattering measurements can be used to investigate compression modes in the atomic nucleus. Due to the scalar-isoscalar ($S=0$, $T=0$) nature  of the $\alpha$ particle (${}^{4}$He nucleus), compression modes are strongly excited in inelastic $\alpha$-scattering experiments at small angles. The use of the $\alpha$-particle probe is a well-established technique that has been extensively employed in the investigation of isoscalar giant resonances in a wide range of nuclei \parencite{PhysRevLett.38.676,HARAKEH1979373,PhysRevLett.42.1121,PhysRevC.21.768,PhysRevC.24.884, PhysRevC.33.1116,PhysRevLett.49.1687,PhysRevC.64.064308,PhysRevC.68.014305,PhysRevC.73.014314,PhysRevLett.99.162503,PATEL2012447}. Usually, the experiments are performed in forward kinematics, where $\alpha$ particles are accelerated to energies ranging from 30 to 100~MeV per nucleon and impinged on a target foil that contains the atomic nuclei of interest. In order to disentangle the different vibration modes (multipolarities, $\Delta L$) excited in the nucleus, measurements of inelastic scattering at very forward angles ($\lesssim 5^\circ$) are required. High-quality magnetic spectrometers allow  the inelastically scattered $\alpha$ particles to be separated from the unreacting beam and other reaction channels. Figure~\ref{fig4} shows a typical spectrum of $\alpha$ inelastic scattering at forward angles on a ${}^{48}$Ca target. The solid angle ($d\Omega$) was integrated over center-of-mass scattering angles around 1.1 and 4.3$^\circ$, respectively. As can be seen, the resonance is concentrated in the energy range of 10 to 30~MeV. There is a significant contribution from the continuum that is observed at higher excitation energies. The solid lines show the continuum chosen for the analysis, which was determined by a parameterization fitted to the high excitation region of the spectra \parencite{PhysRevC.65.034302}.

These experiments were successfully performed in the past in several laboratories around the world. Currently, the research of giant resonances with stable nuclei is mainly developed in facilities such as the Research Center for Nuclear Physics (RCNP) \parencite{rcnp}, Texas A\&M University (TAMU) \parencite{tamu} and iThemba LABS \parencite{ithemba}. In these facilities, the scattered particles are momentum-analysed by high-resolution spectrometers and focused onto the focal-plane detectors. The detector systems are usually composed of a combination of position-sensitive multiwire drift chambers, proportional counters, ionization chambers, and scintillation detectors, which enable the identification of the scattered particles as well as the reconstruction of their trajectories. The ray-tracing technique allows for the reconstruction of the scattering angles at the target location and excitation energies. \par

Many vibration modes can be excited via inelastic scattering measurements. However,  these giant resonance components are usually overlapping  in the experimental spectra.  The procedure for extracting multipole strength distributions from the experimental cross section is known as multipole-decomposition analysis (MDA). In the MDA method, angular distributions are obtained in small   excitation energy bins ($\sim 500$~keV) and fitted with a linear combination of distorted-wave Born approximation (DWBA) distributions for angular-momentum transfers as

\begin{equation}
 \left( \frac{d^2\sigma}{d\Omega dE}\right)^{\text{exp}} = \sum_L a_L(E_x) \left( \frac{d^2\sigma}{d\Omega dE}\right)^{\text{DWBA}},
\end{equation}
where $a_L(E_x)$ is the fraction of energy-weighted sum rule (EWSR) for each multipole ($L$) and the superscripts (exp, DWBA) denote the experimental and theoretical cross sections, respectively.  The theoretical cross sections are obtained assuming 100\% exhaustion of the EWSR for each multipole \parencite{harakeh2001}.  This procedure is justified since the angular distributions of $\alpha$ inelastic scattering for each transferred angular momentum ($\Delta L$) are well characterized by DWBA calculations \parencite{PhysRevC.55.285}. The strength distributions as fractions of the EWSRs for the different multipolarities are extracted from the fitted $a_L(E_x)$ coefficients in the MDA with the following expressions \parencite{harakeh2001}

\begin{equation}
S_0(E_x) = \frac{2\hbar^2 A \langle r^2 \rangle}{m  E_x}a_0(E_x),
\end{equation}

\begin{equation}
S_1(E_x) = \frac{3\hbar^2 A }{32 \pi mE_x}\left(11\langle r^4\rangle -\frac{25}{3}\langle r^2 \rangle^2 -10\epsilon \langle r^2 \rangle  \right)  a_1(E_x),
\end{equation}

\begin{equation}
 S_{L\geq2}(E_x) = \frac{\hbar^2 A }{8\pi m  E_x} L(2L+1)^2 \langle r^{2L-2} \rangle a_L(E_x),
\end{equation}
where $m$, $A$ and $\langle r^N \rangle$ are the nucleon mass, mass number and the $N^{\text{th}}$ moment  of the ground-state density, respectively.
The parameter $\epsilon$ is obtained from the centroid-energy systematics of the ISGMR  and ISGQR (isoscalar giant quadrupole resonance)  distributions \parencite{harakeh2001} as $\epsilon=(4/E_\text{ISGQR}+5/E_\text{ISGMR})\hbar^2/3mA$. 

\begin{figure}[!ht]
\centering
\includegraphics[width=0.9\textwidth]{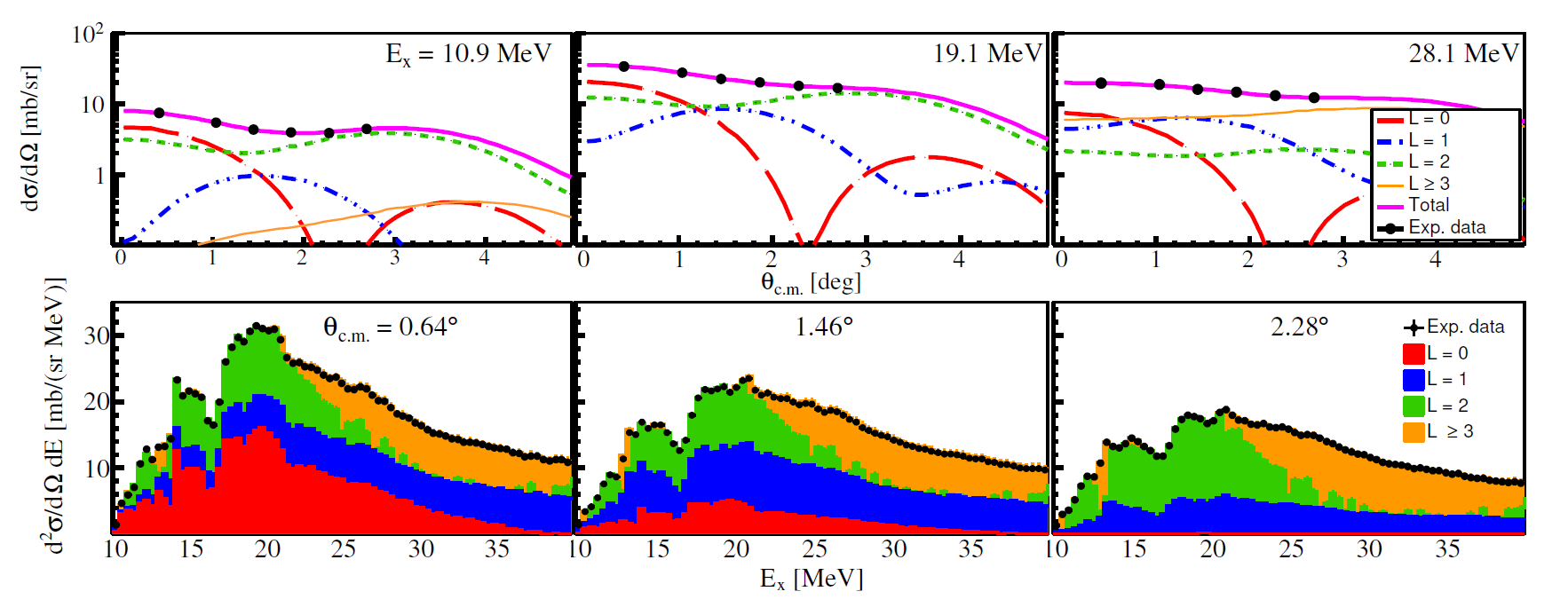}
\caption{\label{fig5} Typical multipole decomposition analysis (MDA) performed for extracting the angular-momentum components of the excitation spectra. The spectra corresponds to inelastic ${}^6$Li scattering on ${}^{24}$Mg. Adapted from \parencite{PhysRevC.104.014607}. }
\end{figure}

Figure~\ref{fig5} shows an example of MDA performed for the analysis of giant resonances in ${}^{24}$Mg excited by ${}^{6}$Li inelastic scattering \parencite{PhysRevC.104.014607}. As can be seen, the giant resonance spectra is composed of various multipole contributions that are disentangled with the MDA. In particular, the monopole strength is the dominant component for the most forward angles. This is evident due to the fact that the $L=0$ angular distribution is maximal at zero degrees, whereas other components become important mostly at large scattering angles and high excitation energies.

The ISGMR is the strongest nuclear compression mode, and its excitation energy is directly related to the nuclear compression modulus. The latter quantity can then be used to derive the nuclear incompressibility of infinite nuclear matter based on microscopic theoretical calculations for the ISGMR strength. The connection between the ISGMR and the incompressibility of nuclear matter  is explained in Section~\ref{sec3}.

\section{Connection between compression modes and the  nuclear-matter incompressibility \label{sec3}}

In the hydrodynamical model, giant resonances can be understood as vibrational modes of a semi-classical liquid drop about its equilibrium shape. In general, these collective oscillations correspond to a superposition of both surface vibration and compression modes. Thus, macroscopic bulk and surface properties of the nucleus can be associated with parameters of the EoS such as incompressibility, surface tension, symmetry energy, etc. In the hydrodynamical model, compression modes are considered to be sound waves in the nuclear fluid. The incompressibility coefficient characterizes the small density fluctuations of the nuclear fluid and its ability to bounce back to its equilibrium shape. In other words, it measures the stiffness of nuclear matter against variations in density. The incompressibility of nuclear matter is defined as \parencite{BLAIZOT1980171}:

\begin{equation}
    K_\text{nm} = 9 \rho_0^2 \frac{d^2 {\cal E}}{d\rho^2}  \bigg\rvert_{\rho=\rho_0},
\end{equation}
where $\rho_0$ is the saturation density and ${\cal E}$ is the total energy per nucleon  introduced in Equation~(\ref{eq1}). This quantity cannot be measured directly, but it can be derived from the strength distribution of a compression mode like the ISGMR. The monopole strength can be assumed to be concentrated in a single peak at energy $E_\text{ISGMR}$, which can be expressed in terms of the incompressibility of finite nuclei $K_A$ as follows
\begin{equation}
    E_\text{ISGMR} = \sqrt{\frac{\hbar^2  K_A}{m \langle r^2 \rangle }},
    \label{eq9}
\end{equation}
where $m$,  and $\langle r^2 \rangle$ are the nucleon mass and mean-square nuclear radius, respectively.  $E_\text{ISGMR}$ is the centroid energy of the ISGMR, and it can be extracted from the experimental monopole strength. Usually, $E_\text{ISGMR}$ is derived from the ratio of different moments of the distribution. For example, the ISGMR centroid energy can be calculated as $\sqrt{m_1/m_{-1}}$ (in the hydrodynamical model) or $\sqrt{(m_3/m_1)}$ (in the generalized scaling model), where $m_k$ is the $k$-th moment of the distribution defined as $m_k=\sum_i E_i^k S_0(E_i)$ \parencite{STRINGARI1982232}. However, Equation~(\ref{eq9}) is only valid for medium-heavy nuclei, where the ISGMR is associated with a single peak at $\sim 80 A^{-1/3}$.  In light nuclei, the monopole strength is spread  over a wide range of excitation energies,  and other states with a different microscopic picture from the radial breathing mode can be distributed along the $E0$ distribution. \par

The relation between the   nuclear-matter incompressibility ($K_\text{nm}$)  and the finite nucleus incompressibility ($K_A$) is generally given in terms of the liquid-drop formula expansion \parencite{BLAIZOT1980171}
\begin{equation}
    K_A = K_\text{nm}  + K_\tau \left ( \frac{N-Z}{A} \right)^2 + K_\text{Coul} \frac{Z^2}{A^{4/3}}+ \dotsm
    \label{eq10}
\end{equation}
 where $N$ and $Z$ are the number of neutrons and protons, respectively, and $A=N+Z$. In the past, the coefficients $K_i$ were extracted from  numerical fits of Equation~(\ref{eq10}) for several $K_A$ values and nuclei.

\begin{figure}[!ht]
\centering
\includegraphics[width=0.8\textwidth]{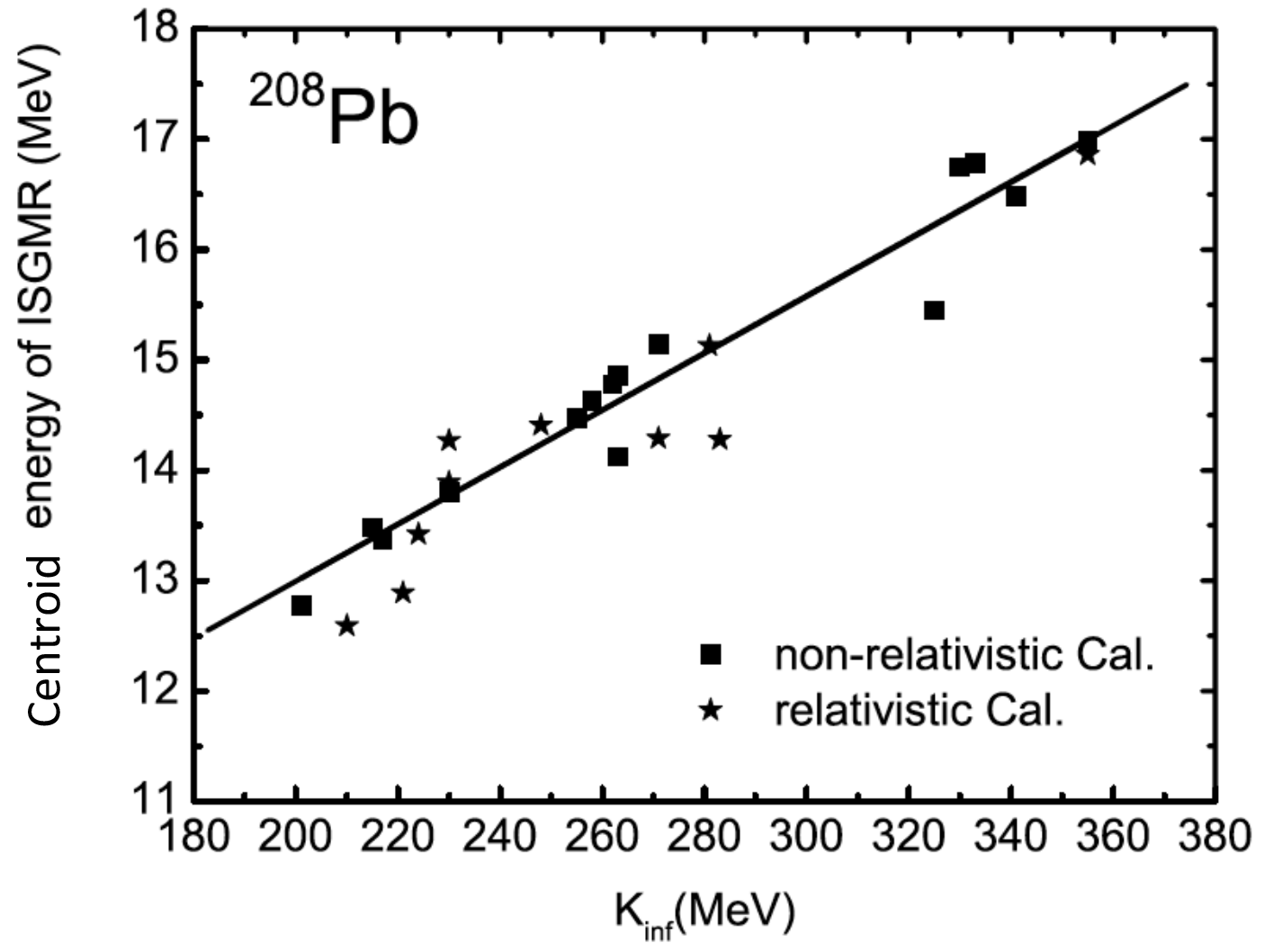}
\caption{\label{fig6} Calculated ISGMR centroid energies of ${}^{208}$Pb as a function of the nuclear-matter incompressibility. Both relativistic and non-relativistic models are included in the calculations using different effective interactions. Adapted from  \parencite{sagawa2019} }
\end{figure}

 However, this procedure may lead to inconsistent results \parencite{PEARSON199112}. The best method for deducing $K_\text{nm}$ relies on microscopic  RPA (random-phase approximation) calculations,  which are well-known for describing successfully  the experimental ISGMR in medium-heavy nuclei. The RPA calculations are based on effective interactions characterized by different $K_\text{nm}$ values.  The strategy entails constructing a collection of effective interactions that differ in their $K_\text{nm}$ predictions, but still provide an accurate description of the nuclear ground-state properties. Thus, a correlation between $K_\text{nm}$ and $K_A$ (or the ISGMR centroid) is obtained by using RPA calculations for various effective interactions \parencite{BLAIZOT1980171}. As shown in Figure~\ref{fig6} \parencite{sagawa2019}, a linear relationship between these parameters is typically derived from RPA calculations employing a large subset of Skyrme, Gogny and relativistic mean-field models.

 The centroid energy of the ISGMR distribution yields a $K_A$ value that is used to calculate the incompressibility of nuclear matter from the experimental data. This method has been applied to several nuclei, but the most consistent results have been obtained for the closed-shell and  doubly-magic nucleus ${}^{208}$Pb. Relativistic and non-relativistic models have been used to reproduce the ISGMR in ${}^{208}$Pb, and to extract the compression modulus of symmetric nuclear matter.

 While non-relativistic calculations predict a coefficient around  $K_\text{nm} = 220$--235~MeV \parencite{BLAIZOT1995435,PhysRevC.56.3121}, relativistic models give results significantly larger in  the  $K_\text{nm} = 250$--270~MeV range \parencite{PhysRevC.68.024310}. The  discrepancy between relativistic and non-relativistic predictions for nuclear-matter incompressibility resulted in a puzzle for  nuclear theorists. However, it can be explained by the density dependence of the symmetry energy within these models \parencite{PhysRevC.66.034305}. As a result, accurately calibrated theoretical models were built to reproduce simultaneously the ISGMR distribution of   both  ${}^{208}$Pb and  ${}^{90}$Zr nuclei. With this prescription, a consensus has been reached about  the incompressibility coefficient, where the accepted value is $K_\text{nm} = 240\pm 20$~MeV \parencite{PhysRevLett.95.122501,PhysRevC.70.024307,Shlomo2006,Piekarewicz_2010}. In a similar fashion, the incompressibility coefficients of nuclear matter for open-shell nuclei have been determined, but these results indicate lower incompressibility coefficients. This problem is known as the "softness" of open-shell nuclei and is explained in Section~\ref{sec4}.

 \section{Open-shell softness and superfluidity in nuclei \label{sec4}}
 
\begin{figure}[!ht]
\centering
\includegraphics[width=0.8\textwidth]{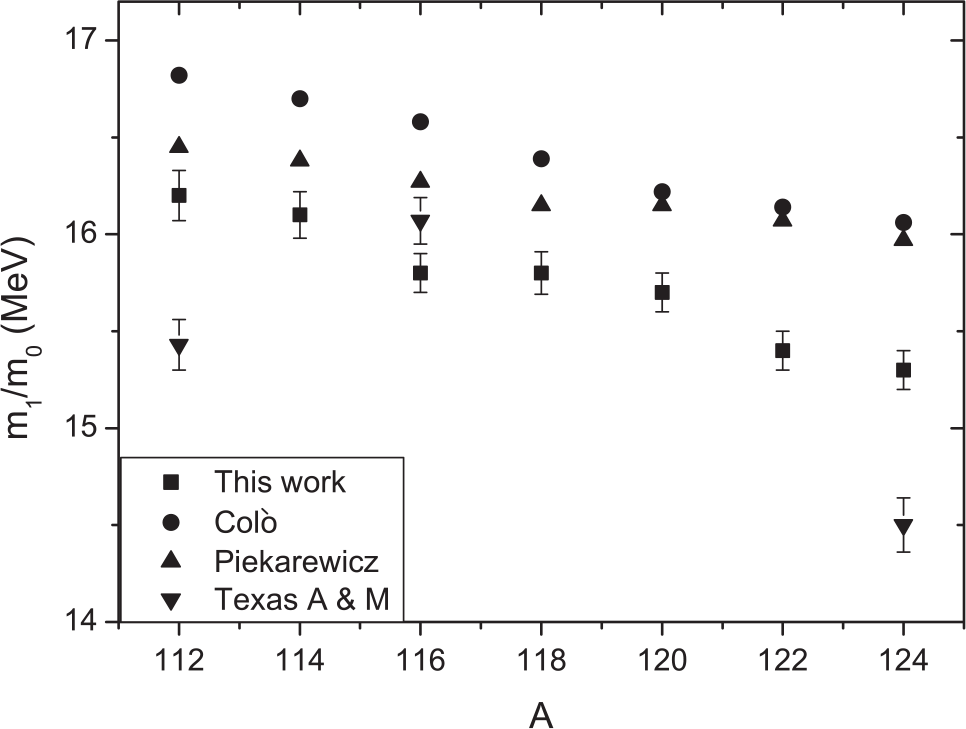}
\caption{\label{fig7}  Systematics of the moment ratios  of the centroid energy for the ISGMR distributions in the tin isotopes. The experimental results (filled squares) are compared with results of non-relativistic RPA calculations (filled circles) and relativistic calculations (triangles). Adapted from \parencite{PhysRevLett.99.162503}. }
\end{figure}

 In recent decades, the ISGMR has been investigated for a wide variety of nuclei across the periodic table.  Most of the theoretical efforts have been devoted  to  develop models that consistently describe "doubly closed shell" nuclei such as ${}^{208}$Pb and ${}^{90}$Zr. Relativistic and non-relativistic models, which successfully describe the ISGMR energies for ${}^{208}$Pb and ${}^{90}$Zr, were also tested with open-shell nuclei. However, large discrepancies were found between the experimental and predicted centroid energies for many nuclei. Experiments with tin isotopes (${}^{112-124}$Sn), for example, revealed that theoretical models consistently overestimate the centroid of the ISGMR distributions by nearly 1~MeV \parencite{PhysRevLett.99.162503}. Figure~\ref{fig7} shows  the centroid energy (given by the moment ratio $m_1/m_0$) of tin isotopes as a function of mass number. As can be seen, results from non-relativistic \parencite{PhysRevLett.99.162503} and relativistic \parencite{PhysRevC.76.031301} models overestimate the experimental data and have a different shape. These findings are surprising given that ${}^{112-124}$Sn  nuclei appear to be "soft"  (lower incompressibility) in comparison to ${}^{208}$Pb and ${}^{90}$Zr. Thus,  experimental data for tin isotopes indicated that the extracted nuclear-matter incompressibility would be significantly lower than the accepted value of $K_\text{nm} = 240\pm 20$~MeV.

Systematic measurements with targets of ${}^{106,110-116}$Cd \parencite{PATEL2012447} and ${}^{94-100}$Mo \parencite{HOWARD2020135608} yielded similar results. These results confirmed the softness of open-shell nuclei with respect to ${}^{208}$Pb and ${}^{90}$Zr. Due to  their proximity to tin isotopes, cadmium results may be expected to have a significant offset from the theoretical models. However, experimental data for around $A=90$ elucidated that the softness effect appears to gradually increase with the addition of nucleons to the ${}^{90}$Zr core. \par

The theoretical ISGMR strength distributions are extremely sensitive to symmetric nuclear matter and symmetry energy (Equation~(\ref{eq1})). For instance, a stiff symmetry energy could explain the softening of nuclear incompressibility \parencite{PhysRevC.79.054311}. Therefore, it is essential to set more precise limits on symmetry-energy parameters such as the slope ($L$), which relates to the pressure at saturation density of pure neutron matter \parencite{PiekarewiczPT2019}. Recent measurements of the neutron-skin thickness in ${}^{208}$Pb  provided a stringent constraint in the slope parameter \parencite{PhysRevLett.126.172502,PhysRevLett.126.172503}. This new value systematically overestimates current experimental and theoretical limits, which results in a larger symmetry energy \parencite{PhysRevLett.126.172503}. Nevertheless, the ''softness`` of open-shell nuclei seems to be related to factors other than a reduction of the symmetry-energy value. In order to fit the ISGMR centroid energies of tin isotopes, effective interactions with a stiff symmetry energy were tried \parencite{Piekarewicz_2010}. However, the puzzle remains unsolved because  the model underestimates the ISGMR centroid energies of ${}^{208}$Pb. \par

\begin{figure}[!ht]
\centering
\includegraphics[width=0.8\textwidth]{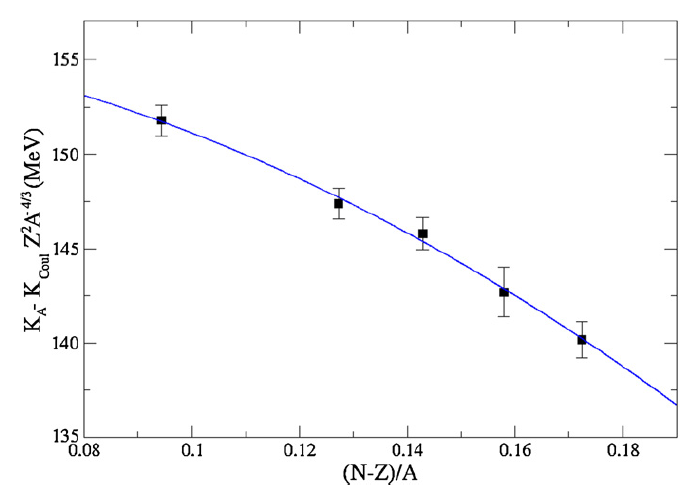}
\caption{\label{fig8}  $K_A - K_\text{Coul} Z^2 A^{-4/3}$ values obtained from experimental ISGMR distributions for cadmium isotopes (${}^{106,110-116}$Cd) as a function of the asymmetry parameter $(N-Z)/A$. The solid line represents a quadratic fit to the data. Adapted from \parencite{PATEL2012447}. }
\end{figure}

Experimental ISGMR distributions along isotopic chains with a wide range of $N/Z$ ratios constrain the incompressibility of symmetric nuclear matter and the asymmetry term, $K_\tau$, from the liquid-drop expansion (Equation~(\ref{eq10})).  $K_\tau$ is related to the neutron-proton asymmetry $(N-Z)/A$ and it is experimentally accessible, but it should never be regarded as the incompressibility coefficient of infinite matter away from the symmetric $N=Z$ limit \parencite{PhysRevC.79.054311}. Thus, from Equation~(\ref{eq10}), $K_A - K_\text{Coul} Z^2 A^{-4/3}$ can be used to extract $K_\tau$ from measurements along isotopic chains covering a wide range in $(N-Z)/A$  ratios. It is worth noting that $K_\text{Coul}$ is well constrained by a set of energy-dependent functionals with a value of $5.2(7)$~MeV \parencite{PhysRevC.76.034327}. However, as previously stated, direct fitting of the coefficients of the liquid-drop formula  (Equation~(\ref{eq10})) does not provide adequate constraints on the value of $K_\text{nm}$. Therefore, this expression should only be used to demonstrate the nearly quadratic relationship between $K_A$ and the neutron-proton asymmetry. Figure~\ref{fig8} shows the quadratic dependence between  $K_A - K_\text{Coul} Z^2 A^{-4/3}$  and $(N-Z)/A$ for cadmium isotopes extracted from Ref.~\parencite{PATEL2012447}. $K_\tau$ was extracted from the fit with the value $-555\pm 75$~MeV, which is also consistent with the value obtained for tin isotopes, $-550\pm 100$~MeV, and  from the analysis of the isotopic transport ratios in medium-energy heavy ion reactions, $-370\pm 120$~MeV \parencite{PhysRevC.80.014322}. \par

\begin{figure}[!ht]
\centering
\includegraphics[width=0.5\textwidth]{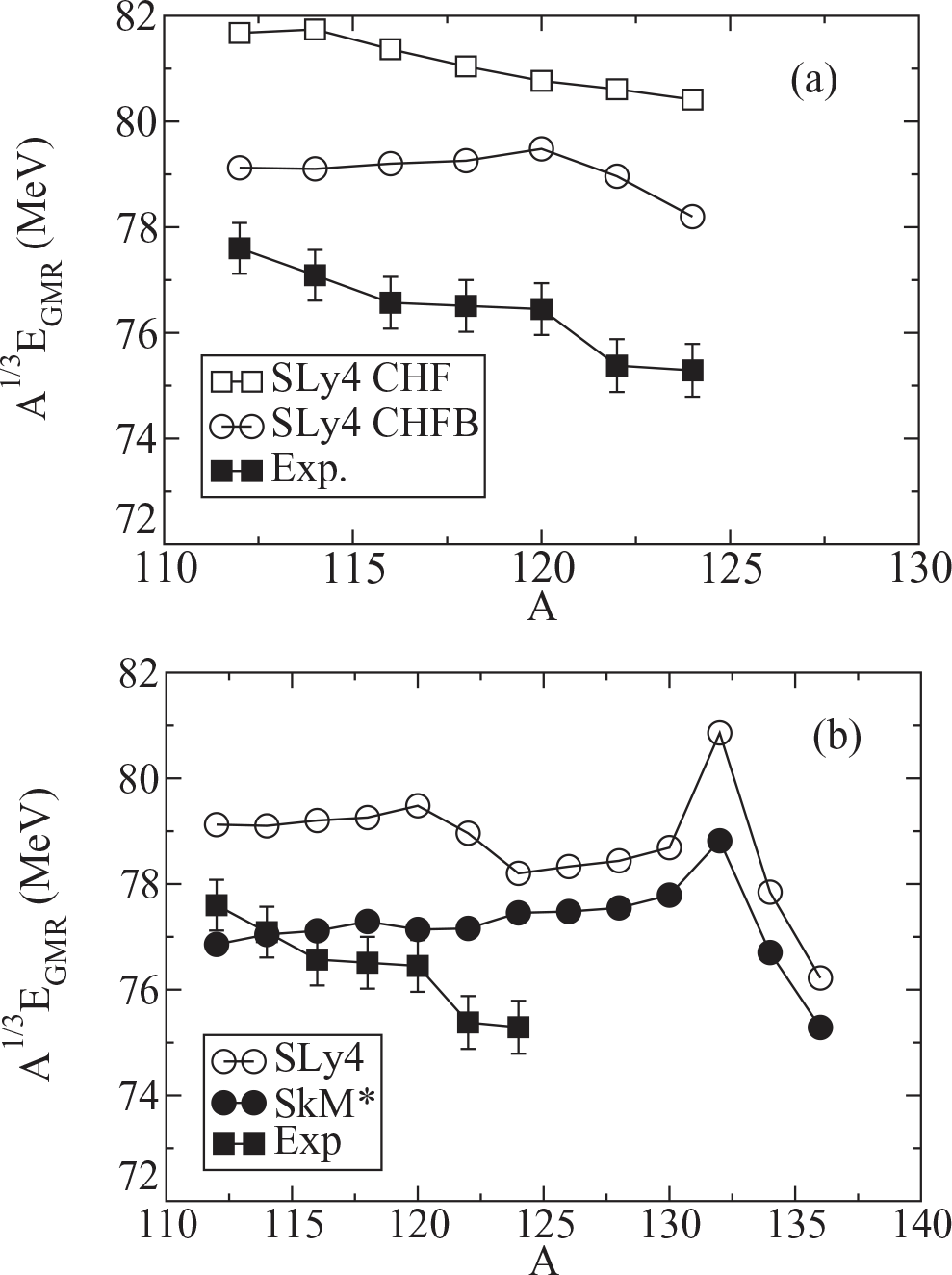}
\caption{\label{fig9} (Top) Centroid energies of the ISGMR distributions in   ${}^{112-124}$Sn (multiplied by $A^{1/3}$) calculated with constrained HF and constrained HFB methods and the SLy4 interaction. (Bottom) Same for ${}^{112-136}$Sn using the  constrained HFB method and the SLy4 and SkM* interactions. Adapted from \parencite{PhysRevC.80.011307}.  }
\end{figure}

Superfluidity effects are predicted to play a significant role in nuclear matter, with important implications for neutron-star phenomena \parencite{MIGDAL1959655,Takatsuka93}. In Fermionic systems, such as the atomic nucleus, superfluidity is induced by the pairing correlation, which binds nucleon-nucleon pairs and substantially reduces the viscosity of the nuclear system. Therefore, superfluidity effects may contribute to "soften" the   nuclear-matter incompressibility in open-shell nuclei.  Significant efforts have been made to investigate the effects of the pairing interaction on the ISGMR strength and the nuclear-matter incompressibility \parencite{PhysRevC.43.2622,PhysRevC.78.064304,PhysRevC.80.011307,PhysRevC.86.054313,PhysRevC.88.044319,Margueron2014}. Self-consistent quasiparticle random-phase approximation (QRPA) calculations on top of a Hartree-Fock-Bogoliubov (HFB) \parencite{ring2004nuclear} approach with pairing interaction were performed to predict the ISGMR strength of a variety of nuclei. Skyrme interactions with different values for the nuclear-matter incompressibility were used to compare the effects on the ISGMR properties. Including the pairing correlation resulted in a consistent reduction of the centroid energies (several hundreds of keV) of the ISGMR distributions in open-shell nuclei. For instance, Figure~\ref{fig9}(a) displays the  centroid energies (multiplied by $A^{1/3}$) for tin isotopes calculated with a constrained Hartree-Fock method with and without  a full Bogoliubov  pairing treatment \parencite{PhysRevC.80.011307}. As can be observed, pairing effects  (constrained HFB) decrease the ISGMR centroids, bringing them closer to the experimental data. However, the  SLy4 interaction ($K_\text{nm}=230$~MeV) used to describe the ${}^{208}$Pb data still overestimates the experimental ISGMR centroids of tin isotopes. Only calculations employing interactions with a lower nuclear-matter incompressibility, such as the SkM* force ($K_\text{nm}=215$~MeV),  can reproduce the experimental data (see Figure~\ref{fig9}(b)). Although these results are very interesting, superfluidity effects seem not to be enough to explain the softness of open-shell nuclei with respect to $^{208}$Pb and $^{90}$Zr.

Recently, Particle-Vibration Coupling (PVC) effects have been included in QRPA calculations based on the HFB framework  in order to reach a unified description of the ISGMR energies in the Ca, Sn and Pb isotopes \parencite{qpvc_2022}.  The authors of this work claim that the coupling of particle-vibration configurations enhances the description of the ISGMR width and reduces the centroid energies for the tin isotopes by a few hundred keV. Eventually, calculations employing the effective interactions  SV-K226 and KDE0 ($K_\text{nm}=226$~MeV and 229~MeV, respectively) can simultaneously describe the ISGMR centroids of ${}^{48}$Ca, ${}^{120}$Sn and ${}^{208}$Pb. Similarly, quasiparticle-vibration coupling (qPVC) calculations, including a finite-range effective meson-nucleon interaction, provided a simultaneous description of the ISGMR centroids for nuclei in the lead, tin, zirconium and nickel mass regions \parencite{PhysRevC.107.L041302}. In general, it was observed that the ISGMR centroids can be naturally fine-tuned by coupling the ISGMR to low-energy phonons, predominantly those with quadrupole character. As a result, the coupling causes spreading and an overall shift of the ISGMR centroids downward relative to their value calculated from relativistic QRPA calculations, which can amount up to 1--2 MeV,  particularly  open-shell nuclei. In the future, theoretical approaches beyond QRPA calculations will further improve the agreement with experimental data in different mass regions.

\section{Deformation effects on the ISGMR distribution}

\begin{figure}[!ht]
\centering
\includegraphics[width=0.8\textwidth]{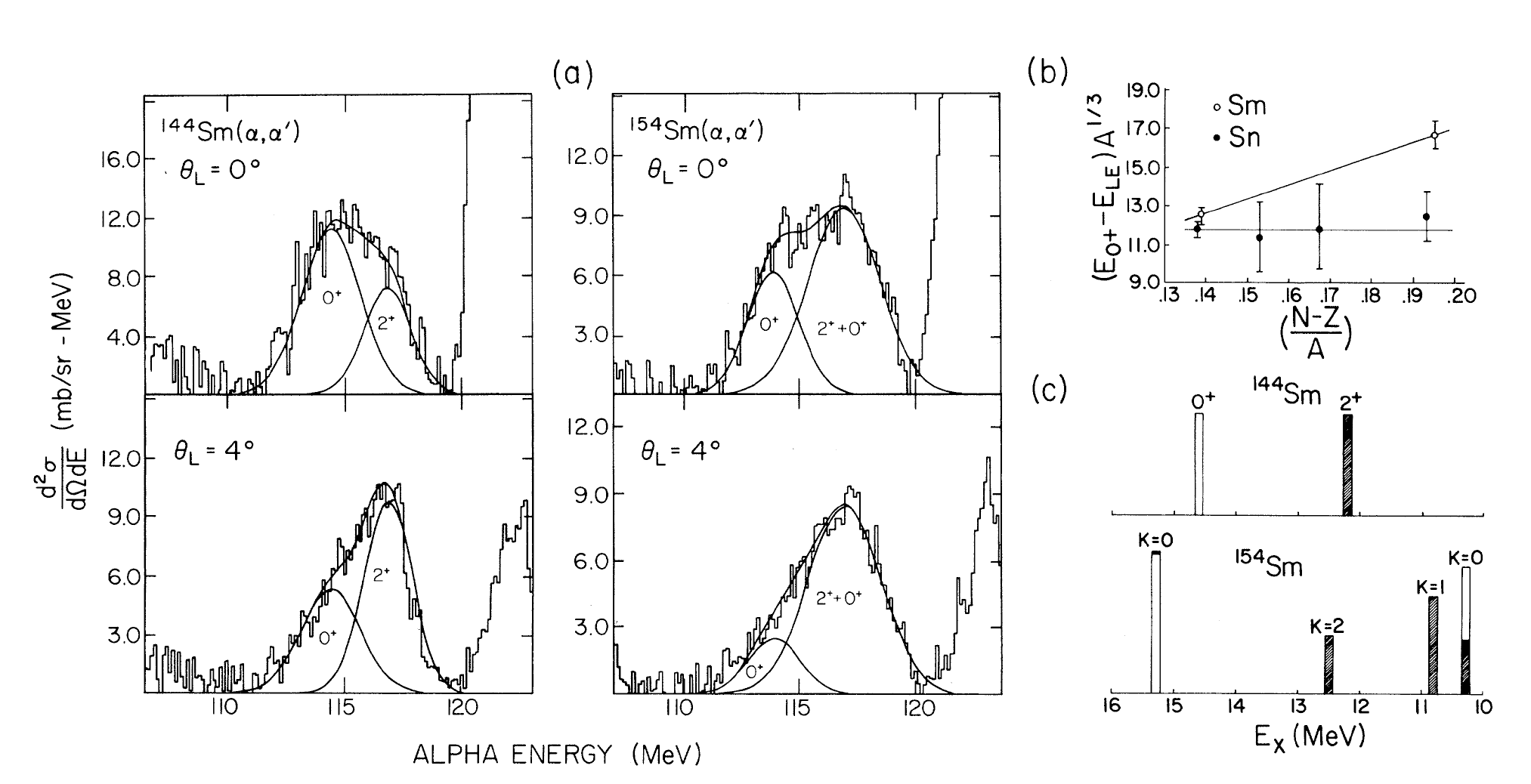}
\caption{\label{fig10} Energy spectra of inelastic $\alpha$ scattering at 0 and $4^\circ$ for ${}^{144}$Sm and ${}^{154}$Sm.  A splitting of the ${}^{154}$Sm spectrum at $0^circ$ is clearly observed. (b) and (c) show the energy difference between the fitted $0^+$ and $2^+$ peaks and their respective theoretical predictions. Adapted from \parencite{PhysRevLett.45.1670}. }
\end{figure}

The nuclear ground-state deformation has a strong effect on the giant resonance strength. For example,  the IVGDR (isovector giant dipole resonance) distribution has a well-known dependence on the nuclear shape, which results in a splitting of the strength distribution. This can be understood as a dipole oscillation with frequencies along the parallel ($K=0$) and perpendicular ($K=1$) axes, which generates a splitting in the IVGDR distribution that is proportional to the nuclear deformation parameter \parencite{harakeh2001}. In a similar fashion, the ISGQR in deformed nuclei is split into $K=0$, 1 and 2 distributions, but the small separation between components results in a broadening of the ISGQR strength \parencite{PhysRevLett.35.552}. The strong $K=0$ coupling between the ISGQR and the ISGMR is the most intriguing effect in deformed nuclei. In particular, the $0^+$ mixing of the monopole and quadrupole strengths leads to a splitting of the ISGMR distribution into two components. This effect has been consistently investigated  in spherical and deformed nuclei, for instance in the isotopic chain  ${}^{144,148,150,152,145}$Sm \parencite{PhysRevLett.45.1670, PhysRevC.68.064602}.  As shown in Figure~\ref{fig10}, the ISGMR strength is split into low and high energy components, and the effect becomes more pronounced for the most deformed nuclei.  In the spherical ${}^{144}$Sm nucleus, the ISGMR distribution is centred around 15~MeV; however, for isotopes with nuclear deformation, part of their strength is shifted to lower energies.

Deformation effects on the ISGMR are difficult to measure in light nuclei since their strength is fragmented over a broad range of excitation energies. In addition, the subtraction of the continuum and instrumental background presents a further challenge for experiments using light nuclei. Experiments  with ${}^{24}$Mg \parencite{GUPTA2015343} and ${}^{28}$Si \parencite{PhysRevC.93.064325} have demonstrated that the ISGMR splitting is also a feature of light nuclei with prolate and oblate deformation. These experimental data are well described by fully consistent axially-symmetric-deformed HFB+QRPA calculations using Skyrme and D1S Gogny interactions \parencite{PhysRevC.77.044313,YOSHIDA10}. \par

\begin{figure}[!ht]
\centering
\includegraphics[width=0.5\textwidth]{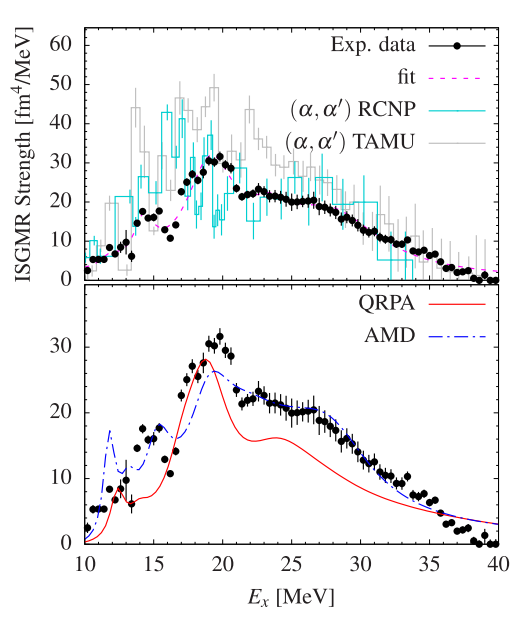}
\caption{\label{fig11} ISGMR strength function of ${}^{24}$Mg. (Top) Inelastic ${}^{6}$Li scattering data  are compared with results from  
$\alpha$-scattering experiments (histograms).  (Bottom) Comparison with theoretical predictions from Antisymmetrized molecular dynamics (AMD)  and fully consistent quasiparticle random-phase approximation (QRPA) calculations.  Adapted from \parencite{PhysRevC.104.014607}. }
\end{figure}

Recently, background-free measurements using the probe $({}^{6}\text{Li},{}^{6}\text{Li}')$ have been performed with a ${}^{24}$Mg target to investigate the effect of the continuum and background subtraction on the ISGMR distribution \parencite{PhysRevC.104.014607}.  Figure~\ref{fig11} compares the ISGMR strength obtained from a $({}^{6}\text{Li},{}^{6}\text{Li}')$ experiment to data from $(\alpha,\alpha')$ measurements with distinct background subtraction.  The $(\alpha,\alpha')$ data from RCNP and TAMU differ primarily in the 15 to 25~MeV energy range. This difference can be attributed to the parameterization employed in the TAMU data for the background subtraction. Unfortunately, this process is model-dependent and may affects the extraction of the giant resonance strength. A two-peak structure can be observed in the RCNP data using both $(\alpha,\alpha')$ and $({}^{6}\text{Li},{}^{6}\text{Li}')$ probes.  This is clear evidence that the ISGMR is splitting as a result of the ${}^{24}$Mg deformation. However, the position of the low energy peak in these experiments differs by 3~MeV. In the future, new background-free measurements with different probes will be important to understand the shape of the ISGMR in ${}^{24}$Mg and other light nuclei.

 \section{Fine structure of the ISGMR}

Extensive research has been conducted in recent years on the fine structure of nuclear collective modes such as the IVGDR \parencite{PhysRevC.89.054322} and the ISGQR \parencite{PhysRevLett.93.122501} by means of high-resolution inelastic scattering experiments.  A quantitative analysis of the fine structure of giant resonances enables the investigation of various mechanisms, such as Landau damping and particle decay, which account for the total width of the resonance.  This type of analysis is based on the wavelet transform \parencite{mallat1999wavelet}, which can be considered an extension of the Fourier analysis that allows  to conserve the correlation between the observable and its transform  \parencite{vonNeumann-Cosel2019}. Usually, the continuous wavelet transform (CWT) is applied to high-resolution energy spectra in order to construct the power spectrum of the signal, including the projection of the absolute values of the wavelet coefficients onto the  scale distribution, which is used to extract lengthlike scales from spectra, i.e., “distances” between features \parencite{PhysRevC.77.024302}. Due to the sensitivity of the wavelet to the level spacing,  it is not possible to directly interpret the wavelet scales as underlying widths. Instead, microscopic calculations of the giant resonance strength, including many decay mechanisms, are also analysed with the CWT and compared to the experiment. The comparison of the experimental and theoretical power spectra allows  the study of  Landau damping in giant resonances and the spreading width via coupling to the continuum or  two-particle–two-hole configurations \parencite{ring2004nuclear}. \par
 
\begin{figure}[!ht]
\centering
\includegraphics[width=0.5\textwidth]{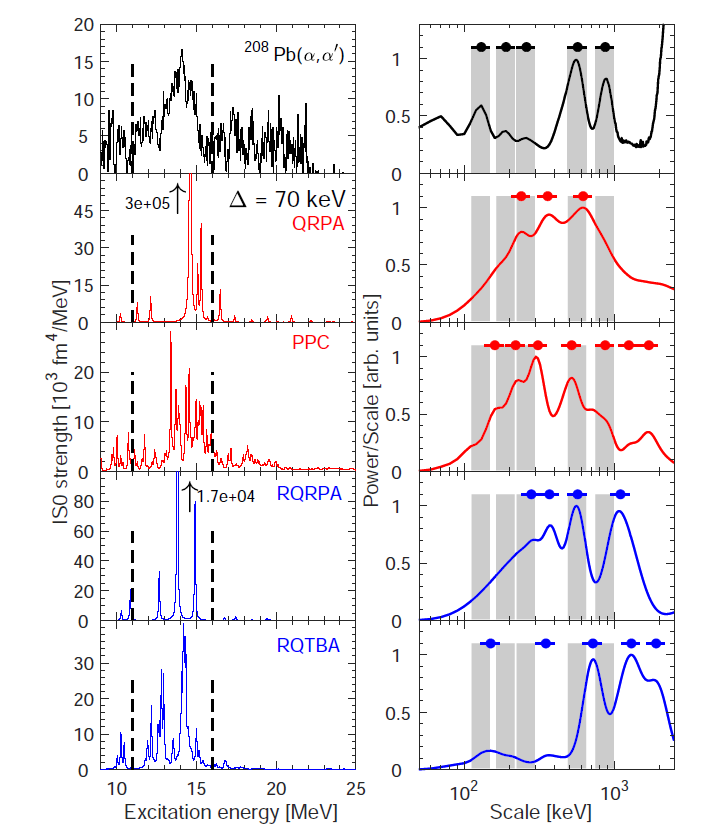}
\caption{\label{fig12}
(Left) Experimental data (top) in comparison with model predictions (rows 2--5) for the ${}^{208}$Pb ISGMR.  The vertical dashed lines indicate the integration region of the wavelet coefficients to determine the power spectra. (Right) Corresponding power spectra. Scales are indicated by filled circles with the associated errors, and for the experimental results additionally by vertical grey bars. Adapted from  \parencite{Bahini_2023}.}
\end{figure}

At present, iThemba LABS is making a significant effort to investigate fine structure effects on the ISGMR for a wide range of nuclei using high-resolution alpha inelastic scattering \parencite{Moodley2019, Bahini2021}. For example, recent results on ${}^{58}$Ni, ${}^{90}$Zr, ${}^{120}$Sn and ${}^{208}$Pb  have been presented in Ref.~\parencite{Bahini_2023}. Figure~\ref{fig12}  compares the power spectra obtained from the experimental data and  phonon-phonon coupling (PPC), QRPA, RQRPA and relativistic quasiparticle time blocking approximation (RQTBA) calculations. Several scale values ranging from several hundred keV to a few MeV are evident in both theoretical and experimental data, as shown in Figure~\ref{fig12} with the filled circles on top. A few experimental energy scales are well reproduced by the theoretical models.   A detailed comparison with additional theoretical calculations, including multiple decay mechanisms, will enable the extraction of important information about the damping of the ISGMR in nuclei. In the future, the technique will be applied to the study of fine-structure effects on a wide range of nuclei.

 \section{Novel techniques for investigating giant resonances in unstable nuclei}

During the last decades, measurements of giant resonances over a broad range of nuclei have been successfully performed in normal kinematics by employing dedicated spectrometers to separate the inelastically scattered particles at small angles in the center-of-mass ($\theta_\text{c.m.}$) system. However, this technique is limited to studies with stable nuclei because of the difficulty of producing targets of short-lived unstable nuclei. With the availability of radioactive beams, novel techniques based on  inverse kinematics have been developed \parencite{STECK2020103811,BAZIN2020103790}.  The advent of high-intensity radioactive beams and novel detection techniques has generated a strong interest in the study of ISGMR in far-from-stability nuclei. \par

An important advantage of carrying out giant resonance experiments in inverse kinematics is that the scattered recoils at small $\theta_\text{c.m.}$ can be measured at relatively large laboratory angles  ($20^\circ \lesssim \theta_\text{c.m.} \lesssim  50^\circ$). This means inelastically scattered particles are kinematically separated from the beam direction, which in turn is quite favourable for measurements at very forward angles in the center-of-mass frame. However, these experiments are constrained by the low kinetic energies of the scattered recoils, which are usually in the order of a few hundreds of keV. In this case, straggling and energy loss in the target material  play a critical role in the recoil detection. The active target and storage-ring techniques are two possible approaches to overcome these limitations while simultaneously increasing the reaction luminosity \parencite{PhysRevLett.100.042501,ZAMORA201616}. \par

\begin{figure}[!ht]
\centering
\includegraphics[width=0.8\textwidth]{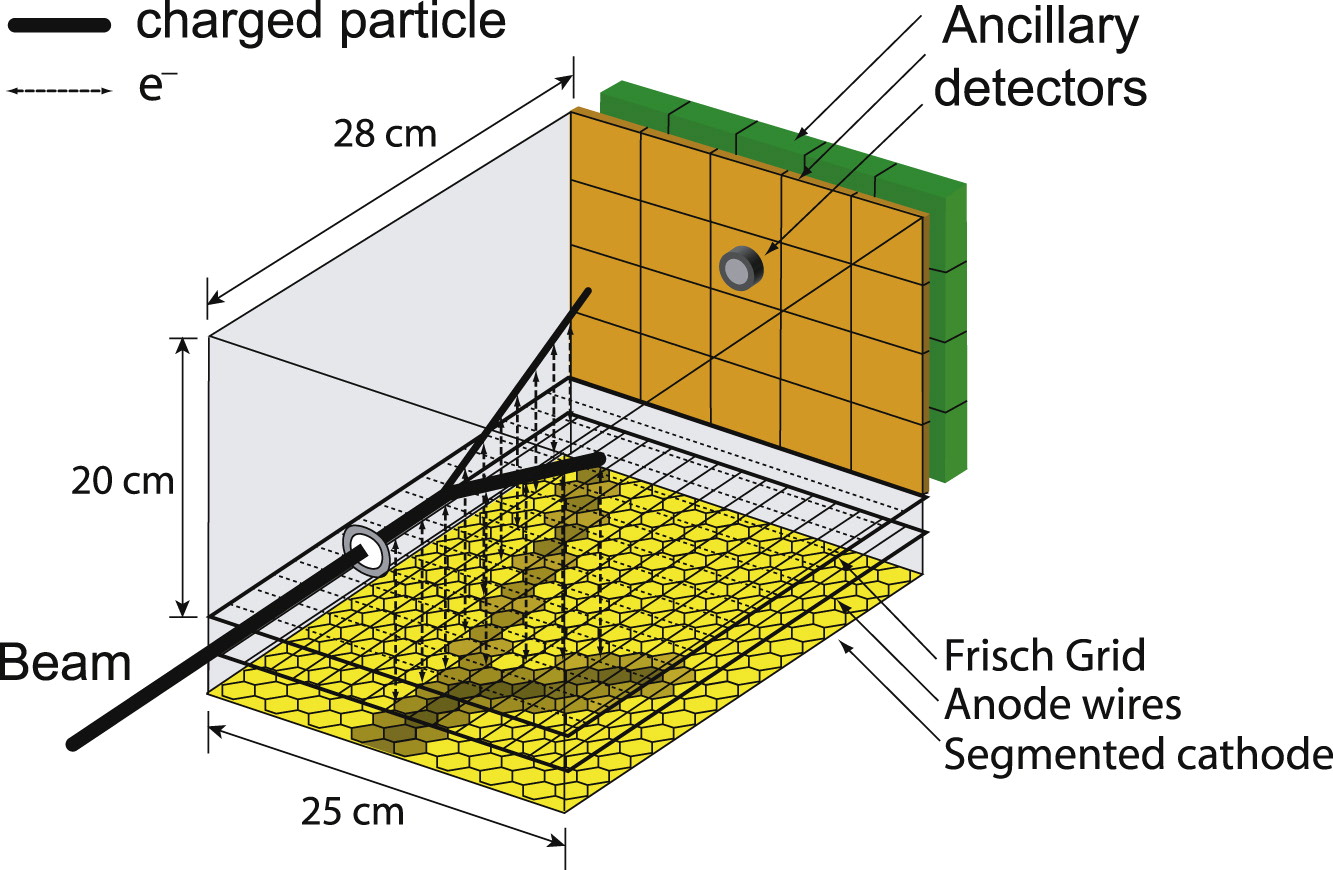}
\caption{\label{fig13}  Schematic representation of the MAYA active target. The beam particles enter  the active volume, inducing a nuclear reaction in the gas. The reaction products may produce enough ionization to induce a pattern in the segmented cathode, after traversing a  Frisch grid \parencite{knoll2010radiation} and a plane of amplification wires. A set of ancillary detectors is used in the exit side of the detector. Adapted from  \parencite{ROGER2011134}. }
\end{figure}

Active target (AT) is a system that combines a target medium and detector in a single device. This detection system enables complete measurements of nuclear reactions with large solid-angle coverage ($\sim 4\pi$) and low-energy detection thresholds. The target thickness can be increased to achieve measurements at high luminosities without any significant impact on the reconstruction of the reaction vertex and angular and energy resolutions. Over the last few years, a new generation of active targets based on the time-projection chamber (TPC) concept has been developed \parencite{BAZIN2020103790}. They enable measurements of energy loss and position hits in 3-dimensional space, which are employed for particle identification and momentum reconstruction. During operation, the beam particles impinge on the TPC active volume and induce nuclear reactions along their path. The reaction products ionize the gas atoms while traversing the active volume and generate electrons. Upon applying a uniform electric field, ionization electrons produced by charged particles along their tracks drift towards a sensor plane at a constant velocity. Typically, the sensor plane is highly segmented ($\sim 10^4$ pads) and allows a precise determination of the energy loss and a reconstruction of a 2D image of the particle trajectory, while the third dimension is extracted from the drift time of the electrons. Figure~\ref{fig13} shows an illustration of the MAYA active target with a reaction event in the TPC volume. The electrons drift perpendicularly to the beam direction, towards the sensor plane at the bottom of the picture. In this setup, ancillary detectors are included at the edge of the TPC for improving particle identification.

\begin{figure}[!ht]
\centering
\includegraphics[width=0.5\textwidth]{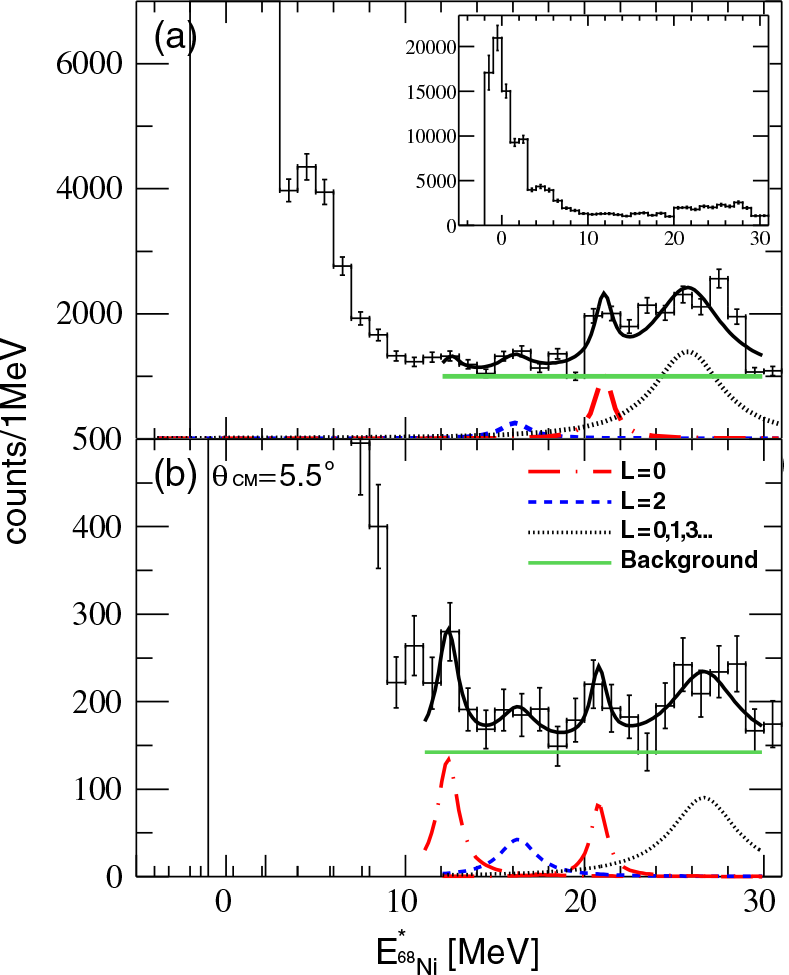}
\caption{\label{fig14}  Excitation-energy spectrum for the reaction ${}^{68}\text{Ni}(\alpha,\alpha')$ measured with the MAYA active target for all angles (top) and $\theta_\text{c.m.}=5.5^\circ$ (bottom).  The subtracted background is indicated by the horizontal green solid line. The data were fit with Lorentzians peaks for the  ISGQR and the ISGMR, respectively. Adapted from \parencite{PhysRevLett.113.032504}. }
\end{figure}

Pioneering measurements of giant resonances with nickel isotopes have been done using the MAYA active-target detector. These experiments were carried out using $^{56}$Ni \parencite{PhysRevLett.100.042501,BAGCHI2015371} and $^{68}$Ni \parencite{PhysRevLett.113.032504} beams, as well as deuterium and helium gas targets. On the one hand, using a deuterium target enabled measurements with a pure gas, but the inelastic scattering cross section at forward angles is considerably smaller relative to alpha-particle scattering. On the other hand, a helium target required an admixture of quench gases, which introduces a considerable background component and makes the accurate determination of the ISGMR strength very challenging. Figure~\ref{fig14} shows the energy spectrum of the reaction ${}^{68}\text{Ni}(\alpha,\alpha')$ measured with MAYA \parencite{PhysRevLett.113.032504} as an example. Several bumps are observed for excitation energies above 10~MeV, which can be attributed to the monopole and quadrupole strengths. Nevertheless, a considerable background component extends over the entire giant resonance region that may affect the extraction of the ISGMR distribution.

Recent experiments with the unstable ${}^{14}\text{O}$ and ${}^{70}\text{Ni}$ beams were successfully performed using both the AT-TPC  (Active Target Time Projection Chamber) \parencite{Bradt2017} and the S800 spectrometer \parencite{Bazin2003}.   This coupled setup enables momentum reconstruction of the beam-like particles and provides reaction-channel tagging as well as an excellent trigger for the acquisition system. An important advance in these experiments was the use of pure helium or deuterium gas as a target and tracking medium. The use of pure gases was facilitated by the use of Multi-layer Thick Gas Electron Multipliers (M-THGEMs), which perform electron pre-amplification before the final-stage avalanche in the micromegas and hence provides higher gain \parencite{Cortesi_2015}. The experiments are still being analysed, but the results will be published soon. In the future, the technique will be extended for ISGMR studies at FRIB \parencite{FRIB2022}  with  high-intensity beams, including  the most exotic nuclei. \par

\begin{figure}[!ht]
\centering
\includegraphics[width=0.8\textwidth]{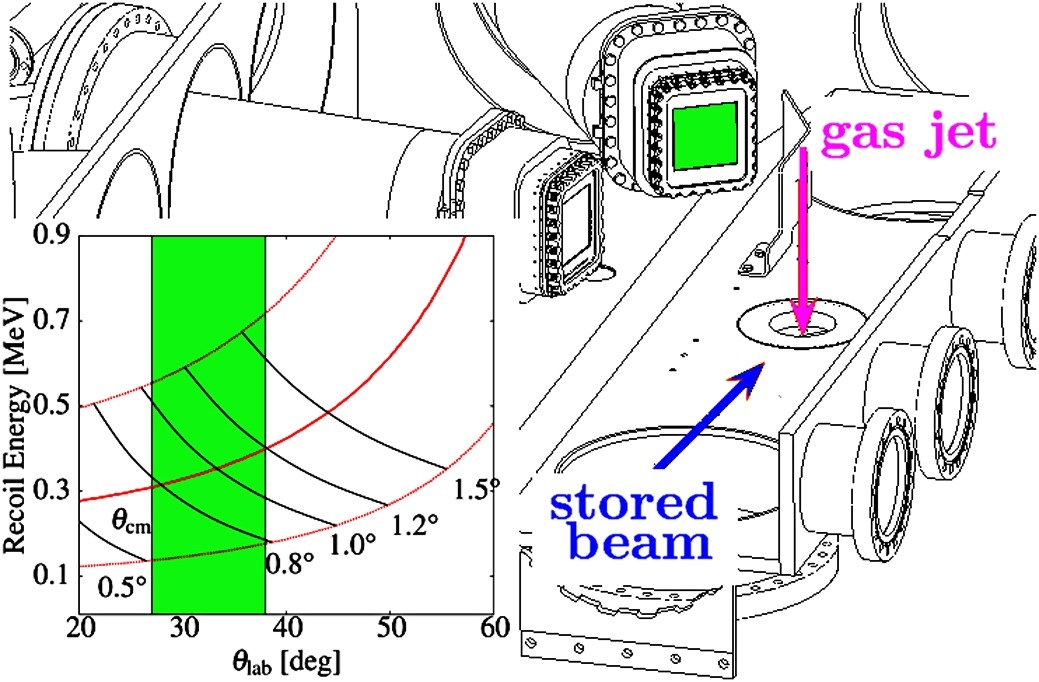}
\caption{\label{fig15} Schematic illustration of the vacuum chamber installed in the storage ring at GSI. The stored beam interacts with the gas-jet target oriented perpendicular to the beam. The detectors were assembled at two internal chambers centered at 80$^\circ$ and 32$^\circ$, with respect to the beam direction. Measurements of isoscalar giant resonances were performed with a detector covering angles from 27$^\circ$ to 37$^\circ$. Kinematics for the excitation of the ISGMR is shown in the inserted plot. Adapted from \parencite{ZAMORA201616}. }
\end{figure}

The storage ring technique is an alternative method to achieve high-resolution measurements of direct reactions in inverse kinematics at very low momentum transfer. The experiments are performed at storage ring devices like the  ESR (experimental heavy-ion storage ring) \parencite{FRANZKE198718} at the  GSI facility. The detection concept is based on the EXL (exotic nuclei studied with light-ion induced reactions in storage rings) project \parencite{fair,1402-4896-2015-T166-014053}, which aims for nuclear structure studies using unstable exotic nuclei in light-ion scattering experiments at intermediate energies and inverse kinematics. The technique employs an internal gas-jet target (e.g., helium) operated at  supersonic speeds with a density in the order of $10^{12}$--$10^{13}$ part./$\text{cm}^2$. This low target density is effectively compensated by the beam revolution frequency ($\sim 10^6$~rev/s), resulting in a significant increase in luminosity.  The detection system is mounted around the point where the gas-jet target and the stored ion beam interact. The experimental setup utilised in Ref.~\parencite{ZAMORA201616} is depicted in Figure~\ref{fig15}.

The detectors are located directly within the ring without any windows. Consequently, the detector system must be compatible with ultra-high vacuum (UHV) and mounted around the internal gas-jet target, which requires a vacuum of $10^{-10}~\text{mbar}$ or lower. In order to achieve such a vacuum condition in the ESR, the chamber must be heated to approximately 150 Celsius for several days prior to the experiment. As shown in Figure~\ref{fig15}, the detector array was installed in a vacuum chamber and consisted of two internal chambers named ''pockets`` covering the laboratory angular ranges of $[74^\circ,88^\circ]$ and $[27^\circ,37^\circ]$ respectively.

\begin{figure}[!ht]
\centering
\includegraphics[width=0.8\textwidth]{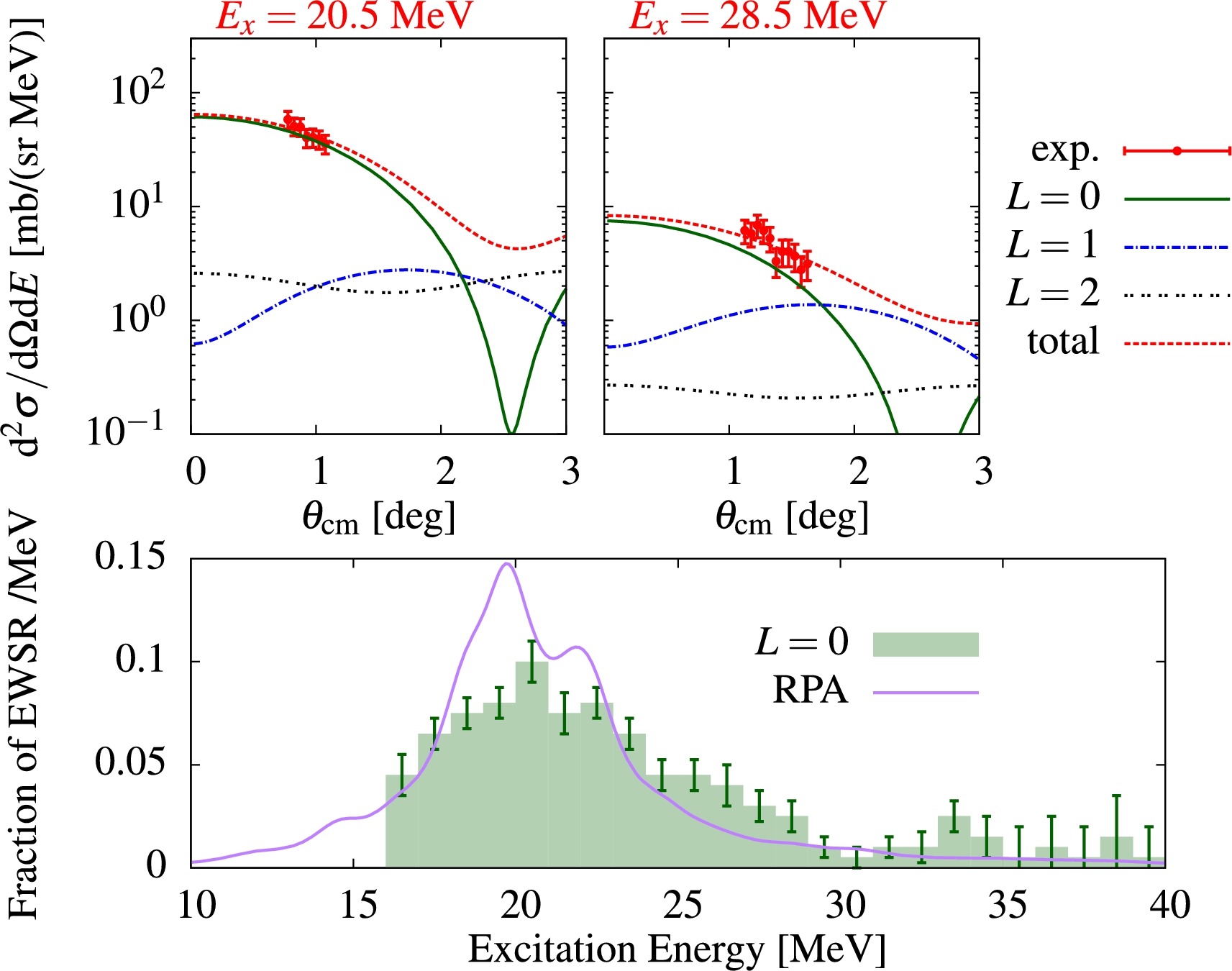}
\caption{\label{fig16} Analysis of the ${}^{58}\text{Ni}(\alpha,\alpha')$ reaction measured at the experimental storage ring in the GSI. The top panels show angular distributions for two measured excitation energies. The bottom figure shows the result of the MDA for the monopole component. The data are compared with a self-consistent RPA calculation presented with a solid line. Adapted from \parencite{ZAMORA201616}. }
\end{figure}

The detector centred at 32$^\circ$ was used to measure low-energy inelastically scattered recoils from the ${}^{58}\text{Ni}(\alpha,\alpha')$ reaction, from which the ISGMR in ${}^{58}\text{Ni}$ can be studied.  This detector measured centre-of-mass angles between 0.5$^\circ$ and 1.5$^\circ$, where the monopole strength predominates. A MDA allowed the extraction of the ISGMR distribution in the energy range from 16 to 40~MeV, as can be seen in Figure~\ref{fig16}. The results are consistent with the analysis of other experiments performed in normal kinematics as well as with theoretical predictions. An  upgraded detector setup will be used in future experiments  to investigate the giant resonances of nuclei that are far from stability.

\section{Outlook}

The ISGMR distribution  is directly related to the incompressibility of nuclear matter, which has significant implications for astrophysical phenomena such as core-collapse supernovae and neutron-star mergers. Despite the fact that the atomic nucleus is many orders of magnitude smaller than a neutron star, the equation of state of nuclear matter can describe both of these (macroscopic and microscopic) systems. Therefore, nuclear reaction experiments provide the means to constrain the models used to understand the behavior of matter in stellar media. These constraints have been fundamental for dedicated astrophysical simulations of core-collapse supernova \parencite{Kuroda_2016}, neutron stars \parencite{Lattimer2021} and compact binary mergers \parencite{Ott_2009}.  \par

Over the past several years, isoscalar giant resonances have been investigated for a large number of nuclei via inelastic-scattering experiments at low-momentum transfer. Theoretical microscopic  calculations are an excellent tool for describing the experimental ISGMR distributions. These models are based on effective interactions characterized by different nuclear-matter incompressibility predictions.

Significant efforts have been devoted to construct  functionals that simultaneously describe experimental data for various nuclei with a certain $K_\text{nm}$ prediction. However,  this has only been possible for ${}^{208}$Pb and ${}^{90}$Zr.   Large discrepancies are obtained for open-shell nuclei, such as tin and cadmium isotopes, for which theoretical models consistently overestimate the centroid energies of the ISGMR, resulting in significantly lower nuclear-matter incompressibility values. A unified description of the ISGMR energies is still an open problem that may require complex coupling configurations beyond pairing and particle vibration.\par

Experimental data along isotopic chains covering a wide range in $N/Z$  ratios, including unstable nuclei, are of paramount importance to determine both the nuclear-matter incompressibility and the symmetry energy more precisely. Novel approaches using inverse kinematics have been developed to achieve giant resonance experiments with unstable nuclei. Pioneering experiments employing active target and storage ring techniques have opened up new opportunities to investigate giant resonances in a large domain of unstable and exotic nuclei in the near future.  The availability of high-intensity radioactive beams in new accelerator facilities will also be essential for obtaining more precise data and providing stringent tests for theoretical models.

\section*{Further Reading}

\begin{enumerate}

 \item Harakeh, M.~N.  \& van~der~Woude, A. (2001). {\it Giant Resonances: Fundamental High-frequency Modes of Nuclear Excitation.} Oxford University Press

 \item Garg, U. \& Col{\`o}, G. (2018). The compression-mode giant resonances and nuclear incompressibility.  {\it Prog. Part. Nucl. Phys. 101}, 55. https://doi.org/10.1016/j.ppnp.2018.03.001

 \item Piekarewicz, J. \& Fattoyev, F. J. (2019). Neutron-rich matter in heaven and on Earth. {\it Physics Today 72(7)}, 30-37. https://doi.org/10.1063/PT.3.4247

 \item Garg, U. (2023). Isoscalar Giant Resonances: Experimental Studies. In: Tanihata, I., Toki, H., Kajino, T. (eds). {\it Handbook of Nuclear Physics.} Springer, Singapore. https://doi.org/10.1007/978-981-19-6345-2\_74

 \item Col{\`o}, G. (2023). Theoretical Methods for Giant Resonances. In: Tanihata, I., Toki, H., Kajino, T. (eds). {\it Handbook of  Nuclear Physics.} Springer, Singapore. https://doi.org/10.1007/978-981-19-6345-2\_72
\end{enumerate}

% % ================================
 %\bibliography{bibliography}  % input bibliography
 \printbibliography
%%%%%====================================================

\end{document}